\documentstyle[preprint,prb,aps]{revtex}

\begin{document}

\draft


\title{Josephson Junction Ladders: Ground State and Relaxation Phenomena.}

\author{Juan J.\ Mazo, Fernando Falo and Luis M.\ Flor\'{\i}a}

\address{Instituto de Ciencia de Materiales de Arag\'on,
Consejo Superior de Investigaciones Cient\'{\i}ficas,\\
Universidad de Zaragoza, 50009 Zaragoza, Spain}

\date{\today}

\maketitle

\begin{abstract}
This paper considers a Josephson Junction array with the geometry of a ladder
and anisotropy in the Josephson couplings. The ground state problem for
the ladder corresponds to the one for the one-dimensional chiral XY model in
a two-fold anisotropy field, which allows of a rigorous characterization of
the ground state phase diagram and the relevant elementary excitations for
the system. The approach to equilibrium, which we study using Langevin
dynamics, shows slow relaxation, typical of systems whose energy landscape
in the configuration space consists of a wealthy of metastable states,
dynamically disconnected.

\end{abstract}


\section{Introduction}

Arrays of Josephson junctions (JJ) in the presence of an external magnetic
field, are probably one of the best examples of physical systems in which
the ideas of competition between interactions (frustration), disorder
effects and complex (glassy) dynamics can be theoretically and experimentally
checked in a controlled way\cite{JJgen}. Most of the effort devoted
in the last decade to these systems has been addressed mainly to the study of
two and three dimensional arrays as models of extreme type II superconductors
in connection with some problems in granular superconductors (as the high
$T_c$ materials)\cite{TCgen}. In this paper we will study a simple geometrical
configuration of links, a ladder, which nevertheless shows an interesting
nontrivial behaviour, and present some results concerning the equilibrium
(ground state) properties as well as the dynamical approach to equilibrium
using Langevin dynamics.

For two-dimensional arrays the ground state is in general unknown for
arbitrary values of the frustration parameter $f$ (which is essentially
the external magnetic field in the appropriate units). Several approaches
have been attempted in order to get close to the truly ground state.
Halsey\cite{H1} proposed a kind of one dimensional solution (Halsey's
staircase) which gives correct configurations but only for certain values
of $f$. He found that this solution provides a highly discontinuous
function for the ground state energy versus $f$, which is in contradiction
with an exact result indebted to Vallat and Beck\cite{V1}. Another way is
a numerical attack of the problem,
by using either phase or vortex variables. Examples of such approach are
the pioneering works of Teitel and Jayaprakash\cite{T1} and more recently,
the "editing method" of Straley and coworkers\cite{S1}. In all these methods
it is assumed that the ground state for rational values of $f$, let say
$f = p/q$, is periodic with periodicity $q$ (or $2q$ in some cases) and
the density of vortices is equal to $f$. However, for the geometry studied
here, because of the free character of the boundary conditions on the upper
and lower branch of the ladder, superconducting currents on them are possible,
allowing of a ground state vortex density different from $f$.

Different studies on superconducting networks with a ladder geometry are
found in literature. Fink and coworkers and Simonin et al. have studied a
ladder of superconducting wires in the Ginzburg-Landau
approximation\cite{Fink}. Current structures commensurate
with the underlying lattice appear as solutions of the Ginzburg-Landau
equations.

Previous to our work, Kardar\cite{K} first and later
Granato\cite{Granato} have study a ladder of JJ in the presence of a
magnetic field and charging effects. Kardar, by doing various approximations
to the interaction potential connects the JJ ladder and the
Frenkel-Kontorova (FK) model with the 'dual Coulomb gas'. Granato
focused his study in the effect of small deviations from a commmensurate
field as a function of the charging energy and in the critical
behaviour in the absence of external field (quantum XY model).

In the remainder of this Introduction, the specific model for the Josephson
Junction ladder will be introduced, along with the notation used and the main
approximations leading to it. In section II, making a judicious choice of the
gauge, it is established the equivalence, regarding the ground state problem,
with the one-dimensional chiral XY model with anisotropy. Going further,
an important argument due to Griffiths and Chou\cite{G} and Sasaki and
Griffiths\cite{SG}, allows the applicability of the vast amount of rigorous
results on the ground states of models of spatially modulated structures
with $convex$ interactions\cite{A} to our system. This provides in a
rigorous way the main properties of the ground state phase diagram,
some of which were already suggested by Kardar and Granato. We also
use the effective potentials method\cite{G} for the explicit computation,
with arbitrary numerical accuracy, of the ground state configuration for
any values of the model parameters.

In section III we perform a linear stability analysis of representative
ground state structures which reveals its persistence as metastable
states outside the domain of $f$ values for which they are ground states.
Many other metastable structures do exist in the system and the
dynamical approach to equilibrium, which we consider using Langevin
dynamics, is characterized by a {\em{constrained dynamics}}\cite{P1} leading
to slow relaxation. This is a remarkable behaviour in the absence of
quenched disorder; i.e., no randomness in the Hamiltonian of the
system\cite{Sethna}.

The system we consider is a ladder of superconducting islands in the
presence of a magnetic field perpendicular to the plane of the ladder
(see Fig.~\ref{ladder}) and we assume that each island is proximity
coupled to its three nearest neighbours. The interaction Hamiltonian
for the system is

\begin{equation}
\label{hjja}
H = - \sum_{<ij>}J_{ij}cos(\theta_{i}-\theta_{j}-A_{ij})
= - \sum_{<ij>}J_{ij}cos\gamma_{ij}
\end{equation}

where $\theta_{i}$ denotes the phase of the superconducting order
parameter at the $ith$ island or site; $\gamma_{ij}$, the gauge
invariant phase difference, is restricted to the interval $(-\pi, \pi]$;
and $A_{ij}$ is proportional to the line integral of the vector
potential $\vec{A}$ between the $ith$ and the $jth$ sites,

\begin{equation}
A_{ij} = \frac{2\pi}{\phi_{0}}\int_{j}^{i} \vec{A}\cdot\vec{dl}
\end{equation}

It is required that  $\sum_{p}A_{ij}=2\pi f$, where $f$ is the ratio
of the flux caused by the external field with the superconducting
magnetic flux quantum and is a measure of the frustration. This relation
expresses the discretized Maxwell equation for the vector potential
and the sum is taken in a clockwise direction over the bonds surrounding
the plaquette $p$ of the lattice. Because the phases are all defined in
the interval $(-\pi, \pi]$, we have $\sum_{p}\gamma_{ij}=2 \pi (n-f)$
where $n$ is the integer that defines the vorticity in each plaquette.
Associated to this value we define the vortex density $\omega$ as the
mean value of $n$ in the ladder.

The hamiltonian (Eq.~\ref{hjja}) is the sum of the Josephson
coupling energies between the neighbouring islands. Here we are
neglecting screening currents by assuming that $A_{ij}$ is fully
determined by the external magnetic field. This assumption is
correct whenever the penetration length is much greater than the island
width. In this situation, there is no flux quantization\cite{Tinkham},
the vortex density being not a flux quanta density, but a fluxoid quanta
density. Also we consider that no charging effects are present.
For the coupling constants, $J_{ij}$, we will assume that  $J_{ij}=J_{x}$
for horizontal links and $J_{ij}=J_{y}$ for vertical ones.

\section{Ground State and Phase Diagram}

	Among the different choices for the gauge one
is particularly convenient: $A_{ij}=+\pi f$ for the upper
links, $A_{ij}=-\pi f$ for the lower ones and $A_{ij}=0$ for vertical
links, which corresponds to a vector potential parallel to the ladder and
taking opposite values on upper and lower branches. Thus

\begin{equation}
\begin{array}{ll}
H=& - \sum_{i}\left[J_{x}cos(\theta_{i}-\theta_{i+1}-\pi f)\right . \\
& \left .+ J_{x}cos(\theta_{i}^{'}-\theta_{i+1}^{'}+\pi f)
+ J_{y}cos(\theta_{i}-\theta_{i}^{'})\right]
\end{array}
\label{hlarge}
\end{equation}

Here $\theta_{i} (\theta_{i}^{'})$ denotes the phase on the
upper (lower) branch of the ladder at the $ith$ step. It is easy to see
that the phase configurations which minimize the
hamiltonian are such that $\theta_{i} + \theta_{i}^{'} = constant$,
independent of $i$; then,
by fixing this constant to $0$ and defining the anisotropy parameter,
$h=J_{y}/2J_{x}$, one obtains the equivalence between the following
hamiltonian,

\begin{equation}
\label{hsimple}
H = - 2J_{x}\sum_{i}\left[cos(\theta_{i}-\theta_{i+1}-\pi f)
+ hcos(2\theta_{i})\right]
\end{equation}

and(Eq.~\ref{hlarge}), regarding the ground state configurations (and other
local minima for the energy).

The hamiltonian (Eq.~\ref{hsimple}) describes a one-dimensional
chiral XY model in a two-fold anisotropy field\cite{B}. It belongs
to a general class of one-dimensional models of spatially modulated
structures\cite{Griffiths}, the simplest of them being the FK
model, extensively studied by Aubry\cite{A}.
The equilibrium properties of these models depend crucially on the
convexity of the interaction potential. In the model defined by
Eq.~\ref{hsimple}, the nearest neighbour interaction potential
is nonconvex. However, it can be proved\cite{G,SG} that only the
convex part of the interaction term plays a role in determining the ground
state configurations, so that the ground state properties are those of a
convex model; of course, the situation may be different if
one considers different aspects of the model, other than the ground state.

The essential physics of the model is the competition between the
anisotropy term (coming from the vertical Josephson couplings) which
tends to pin the phases value to $0$ or $\pi$, and the interaction
term (coming from the horizontal Josephson couplings) which tends to fix
$\theta_{i} - \theta_{i+1} = \pi f$, that is, it tries to keep the value
of the vortex density at the frustration value, $\omega=f$.
The ground state phase configuration at given values of the parameters
($h, f$) is the result of the compromise between both competing
tendencies. In order to compute them
we use the effective potentials method\cite{G} which has become
the standard method to obtain the phase diagram for this type of models.
When used in combination with Newton and relaxation techniques it gives
the ground state configuration with arbitrary accuracy. The computed
phase diagram is shown in Fig.~\ref{diagram} where, for clarity, only
a few transition lines are represented. Characterizing a ground state by
the value of the vortex density ($\omega$), we see in the
Fig.~\ref{staircase}, as predicted by the rigorous arguments mentioned
above\cite{G,SG}, that $\omega(f)$, for fixed $h$, is a Devil's
staircase: a continuous function but such that for each rational
value of the vortex density there is an interval of values of the
field for which $\omega$ remains constant.

This phase diagram is quite different from the one expected for the
isotropic ($h=0.5$) two-dimensional JJ arrays. For the 2D system, though
there is no rigorous proof for it, it is assumed that there are no
intervals of stability for rational values of $\omega$, that is,
$\omega=f$ everywhere\cite{H1,T1}. In the case of the ladder, however,
the vortex density is not equal to the field. If we move along inside
a step, the ground state configuration for the gauge invariant phase
differences do change with $f$, while the vortex configuration remains
unchanged; correspondingly, supercurrents along the links of the ladder
keep varying to compensate for the increase of the field with no change
in the vortex (fluxoid) density. In the case of the ground state for zero
vortex density, it is tempting to speak of this effect as a sort of
Meissner effect, but one should notice that no flux is expeled from
the ladder, and one cannot interpret the value of $f$ for which the ground
state changes as an analog of the critical field for superconductors: it is
the fluxoid density what changes from zero at that value, not the flux density.

Although $\omega(f)$ shows such complex aspect the
ground state energy is a continuos function of the frustration
(Fig.~\ref{staircase}). This is also the case for a 2D JJ array,
as proved analytically in Ref.\cite{V1}.

The vortex configuration {$n_i$} corresponding to a ground state of vortex
density $\omega$($<1$) is explicitly given by

\begin{equation}
n_i = \chi_{\omega}(i\omega +\alpha)
\end{equation}
where $\alpha$ is an arbitrary constant and $\chi_{\omega}(x)
=\chi_{\omega}(x+1)$ is the characteristic function of the interval
$[0,\omega)$:
\begin{equation}
\chi_{\omega}(x)=\left\{
\begin{array}{ll}
1&if\;0\leq x<\omega \\
0&if\;\omega\leq x<1 \\
\end{array}
\right.
\end{equation}

Then, {$n_i$} is a periodic sequence (with minimal period) for rational
values of $\omega$ and a quasiperiodic sequence for irrational values
of the vortex density; it is traditional to speak of commensurate and
incommensurate ground states respectively.

Commensurate ground states are, for any value of the parameter $h$,
pinned and defectible. A configuration is pinned when there exists
a finite value $I_d$ (depinning current) such that if a current
$I<I_d$ is injected into each island on the upper branch and extracted
from each island on the lower branch, the vortex configuration remains
unchanged. In this case the phases change to a new equilibrium
configuration and no voltage appears on the links (the ladder remaining
superconducting). For values of the external current greater than
the depinning current, the phases configuration varies with time and
a voltage can be measured. A defectible configuration admits
discommensurations (defects). An elementary discommensuration in a
commensurate configuration corresponds to a domain wall separating
equivalent vortex configurations which are shifted relative each other,
with minimum increase (or decrease) of the local vortex density
(see figure\ \ref{disc} for an example). Notice that the vortex
configuration of an elementary discommensuration, though only locally
different from the underlying commensurate vortex configuration,
cannot be obtained from this through a finite number of local changes,
but entails the whole rearrangement of a semiinfinite part of the
system. It is important to keep this point in mind when dynamical
approaches to equilibrium are studied. The creation energy of an
elementary  discommensuration goes to zero at the value of the
frustration parameter where the ground state vortex density changes,
and a C-IC transition takes place.

Incommensurate ground states show two different regimes, separated by
an Aubry transition (Transition by Breaking of Analiticity)\cite{A}
at a critical value $h_c$ of the parameter $h$,
which depends on the irrational vortex density, $\omega$. Below this
critical value, the ground state configuration is unpinned (any external
current produces the appearance of voltage on the links) and no defects
can be sustained. In this regime the sequence of gauge invariant phase
differences $\gamma_{ij}$ can be expressed in terms of an analytical
hull function. Thus, in the case of a vertical link
$\gamma_{i} = \theta_{i}-\theta_{i}^{'} = g ( -i2\pi\omega + \beta)$,
with $\beta$ an arbitrary constant (Fig.~\ref{hull}).
The situation changes when $h$ grows above the critical value $h_c$:
the hull function develops infinitely many discontinuities and the
incommensurate ground state becomes there pinned and defectible.
Our estimate of $h_c$ for a golden incommensurability ratio, $\omega =
(3-\sqrt{5})/2$, is $h_c = 0.245...$ . This estimate should certainly
be improved, for we have used rather poor rational approximants of $\omega$.
On the basis of the plausible irrelevance of the deviations from quadratic
of the interaction potential, one can conjecture that the main gap of the
hull function for golden irrational vortex density behaves as $\Delta
\simeq (h-h_c)^{\xi}$, with $\xi = 0.712$, the critical exponent obtained
by Mackay\cite{Mackay} for the Aubry transition in the standard FK model.
The depinning current, $I_{d}$, for the golden incommensurate
structure has been estimated using simulations in the RSJ
approximation\cite{Roma}. It has been shown that $I_{d}$ scales as
$(h-h_{c})^{\nu}$ with $\nu=2.75$ close to the estimation, $\nu=3.011$,
of MacKay for the standard FK model.

\section{Metastability and Relaxation Phenomena}

One of the characteristics of frustrated models is the existence of a large
number of metastable states, a feature which influences dramatically the
dynamical approach to equilibrium. Those states are local minima of the
energy (i.e.: stable solutions of the equilibrium equation, $\frac{\partial H}
{\partial \theta_i}=0$). For the model we are considering, it seems plausible
that the existence of truly chaotic metastable states could be justified, as
an extension of the results obtained in Ref.~\cite{AI_limit}, for the FK
model. In order to illustrate the nature of metastability in the
JJ ladder, we will consider the linear stability of a very restricted
class of configurations, namely those vortex configurations which are ground
state for some values of the model parameters. To analyse the stability of
such states we have worked out the spectrum of small linear perturbations.
The procedure has been the following:

a) First, by fixing the model parameters at values inside the tongue (see
phase diagram) corresponding to the selected particular value of $\omega$,
and using the method of effective potentials, the ground state phase (and
vortex) configuration is obtained.

b) Now we vary finely the parameter $f$ (typically $\Delta f \simeq 10^{-3}$)
and using a Newton method to solve the system of (nonlinear) equilibrium
equations $\frac{\partial H}{\partial \theta_i}=0$, we determine the evolution
of the equilibrium vortex configuration under quasistatic changes in $f$,
along with the energy variation.

c) At each value of $f$, the spectrum of the small perturbations matrix,
$\{ \frac {\partial^{2} H}{\partial \theta_{i} \partial \theta_{j} } \}$,
around the corresponding equilibrium phase configuration, is computed.

In all the cases, a zero eigenvalue is found, which corresponds to the
(continuous symmetry) invariance of the Hamiltonian (\ref{hlarge})
under uniform rotation of all the phases. If the rest of the eigenvalues
are all positive, the configuration is linearly stable, and the state is
a local minimum of the energy for the fixed value of $\omega$ under
consideration; when the lowest eigenvalue takes on a negative value,
the configuration is linearly unstable, usually corresponding to a local
maximum of the energy.

In Fig.~\ref{eigenvalues} we show the evolution of the (non-zero) lowest
eigenvalue of the stability matrix for some simple values of the
commensurability ratio $\omega=1/2,1/3,1/4,1/5,2/5$ when $f$ is varied between
$0$ and $0.5$ and $h=0.5$ ($J_{x}=J_{y}$). Not surprisingly, the range of
stability is much wider than the interval of $f$ values for which a given
state is a ground state: a C-IC transition does not
have associated a lack of stability of the commensurate state (which
remains a local minimum of the energy), but it corresponds to the vanishing
of the creation energy of elementary discommensurations, a feature that
cannot manifest itself through a linear stability analysis. At the edge of
the stability intervals, marked by the negative sign of the lowest
eigenvalue, the extremal character of the configuration changes from a
minimum to a maximum of the energy. An interesting feature is that, at the
border of the stability interval, the gauge invariant phase configuration
is always such that there is a link where the supercurrent reaches its
maximum (critical current) value, which in turn coincides with the
interaction potential leaving its domain of convexity. At this point
any small change in the field cannot be sustained by an increase of the
currents. The vortex structure becomes unstable and the nearer (in
configuration space) stable phases possess a different value
of $\omega$.

In Fig.~\ref{energy} we represent, as a function of the frustration, the
energy of some simple commensurate states. The ground state energy corresponds
to the lower envelope of these (and infinitely many other) curves. The
stability transition points are marked by open circles. We can observe that
near the borders $f=0$ or $f=1/2$ the energies of these commensurate states
are well separated while at intermediate values of $f$ they are very close.
The energies of other (commensurate and incommensurate) stable states, not
included in the figure, also lie around. Besides all those stable states
corresponding to minimum energy configurations which exist as ground
states for some $f$ values, one can construct other metastable states,
corresponding to almost arbitrary rearrangement of vortices, with energies
also lying about. Then, for intermediate values of $f$, the energy landscape
consists of an extremely complex set of local minima, with comparable energies,
corresponding to phase configurations which are, generically, rather separated
in configuration space and, from a dynamical perspective, almost
disconnected, a situation which is sometimes referred to as
{\em constrained dynamics}\cite {P1}. Remind that most of the states
are unreachable from a given one. The dynamically reachable states are
those which come from the annihilation (or creation) of a finite density
of vortices and not from the rearrangement of part of the lattice. This
fact introduces a hierarchy of states in the relaxation dynamics
which is relevant in the glassy properties of the model\cite{Shenoy}.
We can conclude that this model presents ingredients of systems with glass
behaviour: a complex structure of metastables states and constrained dynamics.
It is worthwhile to emphasize that here there is no quenched disorder in the
hamiltonian. As we will see such scenario is strongly confirmed by
numerical simulations of the dynamics in the presence of noise.

The existence of strong dynamical constraints
in configuration space is a microscopic feature which leads to the macroscopic
phenomenon known as {\em slow relaxation}. Such behaviour has been observed in
many systems like spin glass compounds, polymers, granular superconductors,
etc. In order to check this phenomenon in the JJ ladder we study the Langevin
relaxational dynamics\cite{HH}

\begin{equation}
\label{langevin}
\dot{\theta}_{i}(t)=-\Gamma\frac{\partial H}{\partial
\theta_{i}}+\lambda_{i}(t)
\end{equation}
where $H$ is defined by Eq.~\ref{hlarge} and we use $\Gamma=1$ and  $J_{x}=1$;
$\lambda_{i}$ is an additive thermal noise in the phases and satisfies
$\langle \lambda_{i}(t) \rangle=0$, $\langle
\lambda_{i}(t)\lambda_{j}(t^{'})\rangle=2T\delta_{ij}\delta(t-t^{'})$.

We look for the relaxation of commensurate as well as random vortex
configurations for different values of the temperature T. For each one we
have computed the density of vortices as a function of time and followed
the time evolution of the corresponding vortex spatial configuration. In these
simulations we use ladders with $400$, $2000$ and $5000$ plaquettes to avoid
finite size effects. The results presented here have been obtained for
$f=0.25$ and $h =0.5$ ($J_y=J_x$). The ground state for those parameter values
has zero vortex density, $\omega=0$, and the configuration space shows there
an extremely complex structure of metastable states with close energy values.

In the case of an initially ordered configuration, three temperature regimes
are found (Fig.~\ref{ordered}). At very low values of temperature the state,
of course, remains as a metastable configuration. From some
higher value of $T$, which depends on the particular initial state, and
until $T\simeq 0.05$ one observes the decay from the initial ordered
state to a disordered metastable vortex state. Such state is basically
of the same type obtained after the $T=0$ relaxation of a random
phase configuration. For values of temperature above $T=0.05$ the states
relax slowly to the zero vortex ground state configuration. Such relaxation
for the values of the parameters chosen is observable in the range of
temperature from $T\simeq 0.05$ to $T\simeq 0.15$. About this last value of
$T$, thermal activation of vortices and antivortices appears superposed to
the purely relaxational effects.

Many classes of functions have been proposed to fit a slow relaxation curve.
One of the most general is the KWW\cite{KWWref} (Kohlrausch, Williams and
Watts) or two-parameter stretched exponential law, defined by
$\bar{n}(t)\sim exp[-(t/\tau)^{\beta}]$. Slow relaxation corresponds to
values of the parameter $\beta<1$. For $\beta=0$ logarithmic relaxation
is seen. A value of $\beta$ from $0.5$ to $0.7$ is common in glasses. The
fittings of our simulation data give values of $\beta$ between $0.5$ and
$0.9$, depending on temperature (see Fig.~\ref{KWWajustes}). We have not
been able to find a functional dependence of the exponent $\beta$ with
the temperature. However $\tau \simeq exp(\alpha/T)$ as could be
expected\cite{P1}.

Finally, we have also investigated the microscopic characteristics of the
vortex dynamics in the different regimes of the relaxation.
At low temperatures the relaxation of the commensurates phases is
dominated by the nucleation of new structures compatible with the
initial one. The long time state is a metastable one formed by
different commensurate structures separated by domain walls,
see Fig.~\ref{nucleation}. The metastable states reached from random
phase configurations are essentially the same.
At higher temperatures these intermediate metastable states decay to the
ground state configuration. As a result of the characteristics of
configuration space in this range of temperature we find slow relaxation.
Vortices are slowly (thermally) expelled from the array,
see Fig.~\ref{disordered}.

At very high temperatures the dynamics is dominated by the thermal
generation of vortices and antivortices. Phases are randomly distributed
between $0$ and $2\pi$ and $\bar{n}$ approaches to $f$.

\section{Summary and Concluding Remarks}

In this paper we have analized a Josephson Junction ladder in the
presence of a perpendicular magnetic field. The ground state problem of this
system is equivalent to the one of a FK model with convex interparticle
interaction, which allows to apply the Aubry theory for this class of models.
We have calculated exactly the ground state phase diagram which shows tongues
of stability for rational values of the vortex density and a devil's
staircase structure.
Incommensurate structures exist and can be described
by a hull function. Below a
critical value $h_{c}$ the hull function is continuous and the structure is
sliding (zero depinning current). Above such value the hull function
develops discontinuities and a non zero pinning force appears. Modern
microlithographic techniques along with methods
to detect vortices\cite{pan} may check
the above results in large JJ ladders.

Relaxation of arbitrary vortex configurations fits to a "slow dynamics"
function although no structural disorder is present in the model.
Disorder is introduced via initial random configuration and thermal
fluctuations. The existence of a complex structure of multiple
metastable states and a strongly constrained dynamics in the configuration
space are the essential ingredients for this "glassy" dynamics.

The results summarized above have several consequences.
Due to commensurability effects, finite size in the ladder direction
can produce changes in the dynamical response under dc and/or ac
currents\cite{coreanos}. Mismatch in the boundary condition generates
defective vortex configurations with a peculiar dynamics\cite{def}.
This is an important effect in order to interpret correctly
dynamical results in the ladder.
Though the equilibrium properties of the JJ ladder are those of a FK model
with convex interaction, this equivalence does not hold when dealing with
the dynamics of the model.
For instance, the dc driven dynamics in the ladder does not, in general, shows
a unique $V(I)$ as it should occur for a convex model\cite{npr}.
Finally, if screening currents are considered, the anisotropic JJ ladder
can be a good model for long Josephson juntions and stacked JJ\cite{ustinov}.
In this way, an extensive study of the Josephson Junction dynamics
under dc + ac driving currents is in progress.

\section*{Acknowledgments}

	We are indebted to P. J. Mart\'{\i}nez and J. L. Mar\'{\i}n for
many useful discussions on this and related subjects. We thank
to A. V. Ustinov for pointing out Ref. 7. JJM is supported
by a grant from MEC (Spain). Work is supported by project
PB92-0361(DGICYT) and European Union (NETWORK on Nonlinear approach
to Coherent and Fluctuating Processes in Cond. Matt. and Opt. Phys.;
ERBCHRXCT930331)

\begin{figure}
\caption[]{Schematic picture of the JJ ladder. The gauge choice is shown in
the right-most plaquette.
Eq.\ \ref{hlarge} gives the interaction hamiltonian of this system.}
\label{ladder}
\end{figure}

\begin{figure}
\caption[]{The phase diagram of the JJ ladder obtained using the method of
effective potentials. Each phase is defined by the value of $\omega$
and, for clarity, only a few of the transition lines are represented.
The sketchs of the ladders show the vortex configurations of the simplest
commensurate states.}
\label{diagram}
\end{figure}

\begin{figure}
\caption[]{$\omega(f)$ for the ground states configurations when $h=0.2$.
This function is a devil's staircase: a continuous function with an
step for each commensurate value of $\omega$. The inset shows the continuity
of the ground state energy as a function of the frustration.}
\label{staircase}
\end{figure}

\begin{figure}
\caption[]{(a) An elementary discommensuration (DC) in a $\omega=0$ state.
We show the DC from both phase and vortex points of view.
(b1) The two possible vortex sequences for the $\omega=1/2$ ground state.
(b2) Elementary DC in the $\omega=1/2$ commensurate configuration.}
\label{disc}
\end{figure}

\begin{figure}
\caption[]{$2\pi$ modulo representation of the hull function for
the gauge invariant difference of phase in the
vertical links for the $\omega=13/34$ ground state. This state reflects
the behaviour of true incommensurate (irrational $\omega$) phases. Above
a critical value of $h$ the function develops infinite discontinuities
being analitic for values of $h$ lower than the critical one. In this
case $h_{c}\simeq 0.24$.}
\label{hull}
\end{figure}

\begin{figure}
\caption[]{Lowest of the non zero eigenvalues of the matrix of stability
of a state when $f$ is varied. The picture shows the cases of some simple
commensurate states when $h=0.5$. Open circles show transition from a
minimun to a maximun of the energy when $f$ is decreased.}
\label{eigenvalues}
\end{figure}

\begin{figure}
\caption[]{Energy diagram of some commensurate states as a function of $f$.
The lowest envelope is the ground state energy. Open circles mark the limit of
stability of each state, see Fig.\ \ref{eigenvalues}. The dot line is for
$f=0.25$, the value of the field we choose in the relaxation calculations
we present in the Section III. $h=0.5$.}
\label{energy}
\end{figure}

\begin{figure}
\caption[]{Temporal evolution of the mean number of vortices in
the ladder, $\bar{n}$, in the relaxation curves at different temperatures.
The initial state is the $\omega=1/2$ ordered state. This
state is stable at this value of $f$ and $h$ ($f=0.25, h=0.5$).
At low values of $T$
no relaxation is seen. For values of $T$ below $T\simeq 0.05$ the state
decays to a new metastable one. At temperatures above this value it
decays slowly to the $\omega=0$ ground state. At highest temperatures the
thermal generation of vortices is dominant, being $<\bar{n})(t)>_{t}=f$.
In the cases of $T=0.1$ and $T=0.15$ we also show the relaxation
curves of random initial configuration. Those curves are nearly close to
the $\omega=1/2$ relaxation curves.}
\label{ordered}
\end{figure}

\begin{figure}
\caption[]{Slow relaxation curves with random intial conditions have been
adjusted using the two-parameter stretched exponential law in the range
of temperatures between 0.05 and 0.15. Different exponents between
$0.5$ and $0.9$ are found.
($h=0.5,f=0.25$)}
\label{KWWajustes}
\end{figure}

\begin{figure}
\caption[]{Temporal evolution of a vortex configuration in the ladder,
$n_{j}(t)$, showing nucleation processes. Such processes dominate
the 'low temperature' relaxations of ordered structures. Here the
$\bar{n}(t=0)=1/2$ ordered state decays to a disordered metastable state,
see Fig.\ \ref{ordered}. In the picture a black mark represents a vortex.
($h=0.5, f=0.25, T=0.01$)}
\label{nucleation}
\end{figure}

\begin{figure}
\caption[]{Temporal evolution of a vortex configuration in the ladder,
$n_{j}(t)$, showing slow relaxation. Such processes domine
the 'mean temperature' relaxations. Here the initial state is a
$\bar{n}(t=0)=1/2$ ordered state which decays quickly to a disordered
metastable state, see Fig.\ \ref{ordered}, in the manner shown in
Fig.\ \ref{nucleation}. Then, such states decays slowly towards the
corresponding ground state, in this case ($h=0.5,f=0.25, T=0.1$) and
$\omega_{gs}=0$, no one vortex in the ladder.}
\label{disordered}
\end{figure}

\end{document}